\documentclass[aps,amssymb,amsmath, twocolumn, pra, superscriptaddress]{revtex4}
\usepackage{graphicx}
\usepackage{epsfig}
\usepackage{dcolumn}
\usepackage[up]{subfigure}
\usepackage{float}
\usepackage{dsfont}	
\usepackage{relsize}
 \newcommand{\tr}{\textcolor{red}}

\usepackage{epstopdf}

\usepackage{float}
\usepackage{ragged2e}
\usepackage{braket}
\usepackage[font=small,labelfont=bf,justification=raggedright, format=plain]{caption}
\usepackage{setspace}
\usepackage[colorlinks, citecolor = magenta]{hyperref}
\makeatletter
\begin{document}
\title{Bell-inequality in path-entangled single photon and purity of single photon state}
%\title{Purity of single photon source from Bell-inequality in path-entangled state}

\author{K. Muhammed Shafi}
\affiliation{Quantum Optics \& Quantum Information, Department of Instrumentation and Applied Physics, Indian Institute of Science, Bengaluru 560012, India}
\author{R. S. Gayatri}
\affiliation{Quantum Optics \& Quantum Information, Department of Instrumentation and Applied Physics, Indian Institute of Science, Bengaluru 560012, India}
\author{A. Padhye}
\affiliation{Quantum Optics \& Quantum Information, Department of Instrumentation and Applied Physics, Indian Institute of Science, Bengaluru 560012, India}
\author{C. M. Chandrashekar}
\email{chandracm@iisc.ac.in}
\affiliation{Quantum Optics \& Quantum Information, Department of Instrumentation and Applied Physics, Indian Institute of Science, Bengaluru 560012, India}
\affiliation{The Institute of Mathematical Sciences, C. I. T. Campus, Taramani, Chennai 600113, India}
\affiliation{Homi Bhabha National Institute, Training School Complex, Anushakti Nagar, Mumbai 400094, India}

%================================================
%%
\begin{abstract}
Different degrees of freedom of single photons have been entangled and are used as a resource for various quantum technology applications. We present a simple scheme to perform Bell's test and show the violation of CHSH inequality in a path-entangled single photon state using interferometric and its equivalent non-interferometric approach in beam splitter setting. We demonstrate this experimentally by generating and controlling path-entangled state using both, heralded and un-heralded single photons from spontaneous parametric down-conversion.  The experimental results we present show the transition to violation of CHSH inequality when the purity of single photons state visibility increase above 70\% , $\mathcal{P} >  0.7$.  Our procedure using single beam splitter and two detector module for un-heralded single photon source allows a simple way to test for purity of any single photon source and to study quantum correlations on systems driven by dynamics where single particle entanglement with position space is prominent.
\end{abstract}

\maketitle

%============================================================

\section{\label{sec1}Introduction}
%============================================================

Single photons and entangled photon pairs have been a very useful resource for several  experimental tests of fundamentals of quantum physics\,\cite{BT56, KMW95}. Lately, they have been extensively used for quantum technology applications\,\cite{BPM97, MWK96, JSW00, PBW98, MZK12}. Among different approaches, spontaneous parametric down-conversion (SPDC) is one of the matured methods to generate such photons\,\cite{Cou18}. Experimental feasibility to control various degrees of freedom of photons such as polarization\,\cite{KMW95, KFM04, FKW05}, frequency\,\cite{RFP12}, time-bin\,\cite{VRB15} and path\,\cite{KHL17, KEH20, CVO20} independently and simultaneously in different combinations have enabled generation of high-dimensional entangled states\,\cite{EKZ20}. These higher dimensional entangled systems have shown significant improvement in processing quantum information. In contrast to the standard notion of entanglement that has been associated with two or more quantum systems\,\cite{Bel64, CH78, Asp99}, single-particle states are also known to exhibit quantum nonlocality\,\cite{TWC91, Har94,GHZ95, Vai95}.  This was further probed by a series of theoretical\,\cite{BW99, LK00, BJS01, Enk05, DV07, CD08, BCB13, AKB21, Aie21} and experimental investigations\,\cite{LSP02, SLM02, BRL03, BBL04, BAL04, HUH04, CDL08} establishing  the single-particle entanglement  to be between the spatial modes, rather than between the quantum systems. Over time, single-photon entangled states have also been widely used as a resource for quantum information processing\,\cite{SP02, LLC03, SLS10, SSR11, CAA20}  and quantum computation\,\cite{KLM01, SCS21}. 

For more than four decades since the proposal for Bell's test\,\cite{Bel64} and experimental violation of the Clauser, Horne, Shimony, and Holt (CHSH) inequality\,\cite{CHS69}, innumerable experiments have been performed on various systems to establish quantum nonlocality. Characterization of entanglement between systems and between different degrees of freedom of a single particle have also been carried out.  CHSH inequality violation with single photons entangled in momentum and polarization was shown using  a Mach-Zehnder interferometer setting\,\cite{GGZ08, PLM20}. Violation with photon entangled in polarization and orbital angular momentum\,\cite{VAH14} and using homodyne detection measurement scheme\,\cite{LPK17} has also been reported.

In this work we first present a theoretical scheme to show the violation of CHSH inequality in path-entangled single photon state using all four distinguishable spatial modes of the beam splitter in an interferometric setting with local and independent operations on each pair of modes. We will also show its equivalence to non-interferometric setting while using only two distinguishable spatial modes with combined operation  on both the spatial modes and the way it can be realized using polarization degree of freedom of a single photons. Both the scheme involves the use of beam splitters to control the  probability amplitude of presence and absence of photon along the paths and to obtain  the probabilities associated with the four basis states needed to calculate CHSH parameter. We also model decoherence in the form of depolarizing channel\,\cite{SE11}  and by introducing the presence of multi-photons in the source  and show the transition from violation of CHSH inequality to its validity (thermal state) with decrease in purity of single photon state (visibility).  
We demonstrate this experimentally by employing a setup composing of a probabilistic single photon source from the  SPDC using type-II Beta Barium Borate (BBO) crystal. The experiment was performed using both, heralded single photons source and un-heralded single photons source. In the heralded single photons source setting, coincidence counts were used to calculate probabilities of all basis states using five single photon detector module. For un-heralded single photon source, a simplified version of one beam splitter and two detector module was used to show CHSH violation. Identical values of CHSH parameter $S$ was reproduced using both the schemes establishing the theoretically shown equivalence. Using combination of wave plates we have experimentally reduced the entanglement visibility and mimicked the reduction in purity of single photons (depolarizing channel) to show the transition from violation of of CHSH inequality to validity with decrease in purity.  Unlike the previous experimental reports of such results shown interferometer setting\,\cite{GGZ08, PLM20}, our measurement procedure allows a sequence of measurements without the need for any interferometry approach and reduces the experimental resources needed to show CHSH violation in path-entangled photons.  For any quantum communication and computation applications of single photons, purity of source is a key factor. A simple test for purity using CHSH inequality violation presented here will be a very useful resource and will also be useful to experimentally verify and study quantum correlations in systems where single particle dynamics are associated with spatial modes. 

The paper is organized as follows. In Section\,\ref{Analytics} we present an analytical description of a path-entangled single photon in a beam splitter setup and different procedure to calculate CHSH parameter.  After showing the violation of CHSH inequality and equivalence of different approaches discussed, we present a scheme to introduce tunability in splitting of photons along different paths using polarization degree of freedom as a control parameter.  In Section\,\ref{Exptsch}  we present an experimental setup, the measurement procedure and the results showing the violation of  CHSH inequality. We also present the transition from violation to satisfying inequality with decrease in purity of single photon state by using the tunable parameter in the experimental setup. In Section\,\ref{conc} we conclude with our remarks.

%====================================== 
\section{Theoretical Description}
\label{Analytics}
%======================================
\subsection{Bell's-test for path-entangled single photon}
\label{belltest}
%=====================================

\noindent
{\it Bell's test :} Entangled photon pairs in polarization degree of freedom is given by,
\begin{align}\label{two-photonE}
|\Psi\rangle_{EP}  = \frac{1}{\sqrt 2} \Big [ |H\rangle_{1} |V\rangle_{2} +  |V\rangle_{1} |H\rangle_{2} \Big ],
\end{align}
where $|H\rangle$ and $|V\rangle$ represent the polarization states and subscript represent the two photons.
Standard Bell's test procedure on such entangled pairs involve rotation of polarization angle by $\theta$ and $\delta$ on each photon (two associated Hilbert spaces) independently just before they reach spatially separated detectors.   Rotation of polarization angle results in spanning over all four basis states of the two photons. Then a series of coincidence events of detection in the form of $E(\theta, \delta)$ are recorded for different configurations of angles. Using those values, CHSH parameter 
\begin{align}
S = \lvert E (\theta, \delta) - E(\theta, \delta^{\prime}) \rvert  + \lvert E (\theta^{\prime}, \delta) + E(\theta^{\prime}, \delta^{\prime}) \rvert
\label{bellpara}
\end{align}
is calculated. If the value of $S > 2$, violation of CHSH inequality ($S \leq 2$) is observed suggesting nonlocal effect, an entangled state. Theoretically, 
\begin{align}
E(\theta, \delta) = P_{00} + P_{11} -P_{01} - P_{10}
\end{align}
where $P_{ij}$ are the probabilities of different basis states of the composite system. Experimentally they can be obtained from coincidence counts of occurrence in  different combination of basis states. \\
%=================================================================
%=================================================================
\noindent
{\it Path-entangled single photon :}  Path-entangled single photon state is generated when $m$ spatial modes share a single photon. Then the state in $m$ spatial mode will exists in a Hilbert space $\mathcal{H} = \bigotimes_{i=1}^{m} \mathcal{H}_{i}$, where $\mathcal{H}_{i}$ is the Hilbert space of the $i^{th}$ spatial mode. Each $\mathcal{H}_{i}$ will be spanned by  $\{ \ket{0}_{i}, \ket{1}_{i} \}$ representing the photon occupancy, absence and presence of photon, respectively  in the mode. 
%=====================================
\begin{figure}[h!]
  \includegraphics[width=0.48\textwidth]{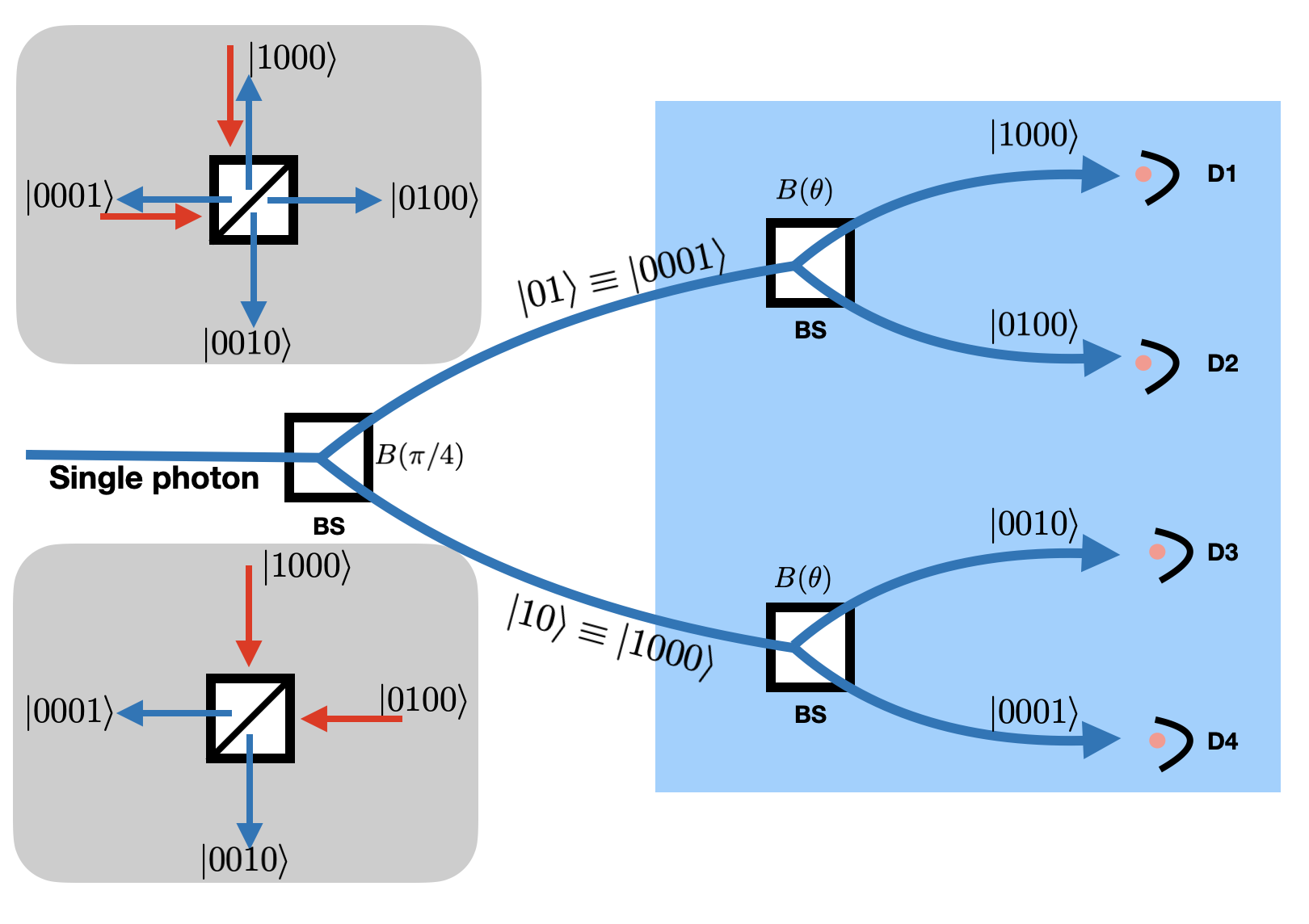}
    \caption{Schematic representation of the input and output modes of the beam splitter.  Two input modes of the beam splitter to realize four and two output modes as basis states is presented in the inset. Physical realization of photon in all four basis states using combination of beam splitters is presented in the main schematic. }  
  \label{Schematic01}
\end{figure}
%=====================================

For a single photon passing through a 50:50 beam splitter, the path-entangled state will be in the form,
\begin{align}\label{path-path}
|\Psi\rangle_{1-2}  = \frac{1}{\sqrt 2} \Big [ |1\rangle_{1} |0\rangle_{2} +  |0\rangle_{1} |1\rangle_{2} \Big ],
\end{align}
where the subscript $1$ and $2$ represent the two spatial modes or paths for the photon.  To calculate CHSH parameter  one has to span over all the basis states associated with  the two spatial mode and photon occupancy. In two photon system described earlier using polarization state  we can perform rotation on the polarization state of the photon and obtain non-zero probability of state $|HH\rangle$ and $|VV\rangle$.  But in single photon occupancy description in two spatial mode, the states $|0\rangle_{1}|0\rangle_{2}$ and $|1\rangle_{1}|1\rangle_{2}$  will always have a zero probability amplitude. However, using single photon state along two different input modes all the four spatial modes of the beam splitter in occupancy representation with non-zero probability can be realized. The basis states in the Fock state representation can then be written as $| n_1, n_2, n_3, n_4 \rangle$  and for a single photon case, $n_1 + n_2 + n_3 + n_4 = 1$.  This can be experimentally replicated  by adding two beam splitters along the two output modes of the first beam splitter as show in Fig.\,\ref{Schematic01}.  Though we have shown identical beam splitter, $B(\theta)$ in the schematic, they need not be identical. In the insets of Fig.\,\ref{Schematic01} we also show the schematic of the beam splitters' spatial modes for single photon and combination of input states for realisation of four and two basis states, respectively. The generic output  state in the photon occupancy representation will be,
\begin{align}
\label{fockstate}
|\Psi\rangle_{1-4}  = a |1000\rangle + b |0100\rangle + c |0010\rangle + d  |0001\rangle.
 \end{align}
When only two of the four output modes have non-zero probability, eliminating the redundant modes, the state will be identical to path-entangled single photon state given in Eq.\,\eqref{path-path}.  
% The same can also be written as four column vector with each element representing each path. 

\begin{figure}[h!]
  \includegraphics[width=0.5\textwidth]{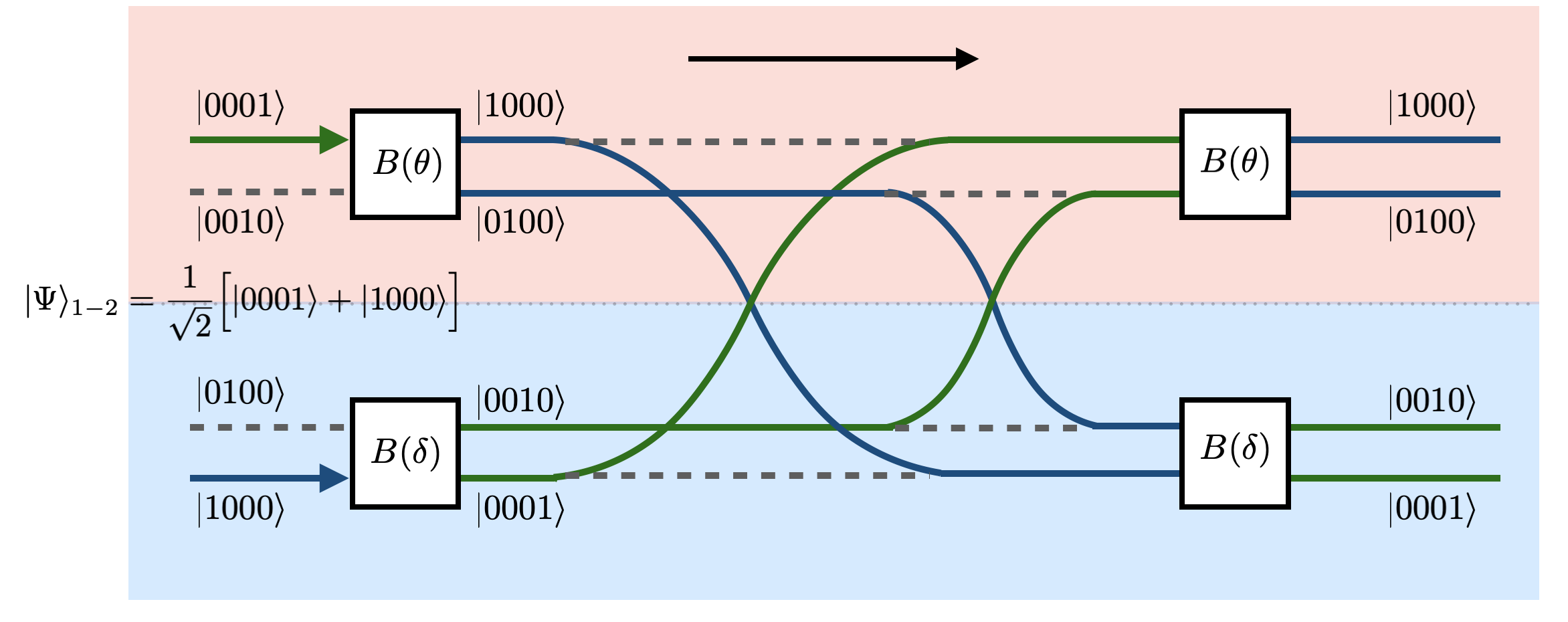}
    \caption{Schematic representation of two input of the beam splitter to realize four output modes when passed through a pair of beam splitter independently on same two modes twice.  Due to change of modes when photon is passed through the first pair of beam splitter, interference of modes is induced in the dynamics when it passes through the second pair of beam splitter. The final output probability of each mode will depend on both the parameters $\theta$ and $\delta$.  Using different combination of beam splitter we can calculate CHSH parameter and show the violation of the inequality.}  
  \label{Schematic01a}
\end{figure}

%===========================================
\noindent
{\it Bell's test on path-entangled single photon using four spatial mode setting:} To perform Bell's test on photon in path-entangled state given by Eq.\,\eqref{path-path}, 
\begin{align}
|\Psi\rangle_{1-2} = \frac{1}{\sqrt 2} \big [ |1 0 \rangle +   |0 1\rangle \big ]  \equiv \frac{1}{\sqrt 2} \big [ |1000\rangle +   |0001\rangle \big ],
\end{align}
we need to independently control the spatial modes using parameter $\theta$ and $\delta$, and calculate $E(\theta, \delta)$ for different combinations.  Therefore, we will set the beam splitter operation of the form,
\begin{align}
B(\theta) = \begin{bmatrix}
	~\cos(\theta) & ~~i\sin(\theta) \\
	i\sin(\theta) & ~~\cos(\theta)
	\end{bmatrix}
\end{align}	
to act only on the input modes  $|0001\rangle$ and  $|0010\rangle$. Similarly, we will set the beam splitter operator  $B(\delta)$ to act only on the input modes $|1000\rangle$ and $|0100\rangle$.  This will ensure that the two pairs of modes are independently controlled. If the measurement probability along the four output modes are dependent on both the parameters, $\theta$ and $\delta$ in spite of independent control,  we can demonstrate violation of CHSH inequality.  In Fig.\,\ref{Schematic01a}, schematic illustration to perform Bell's test and calculate CHSH parameter is given.
  The scheme comprises of action of two independent beam splitter operation with different splitting ratio, $B(\theta)$ and $B(\delta)$ on each pair of input modes.  To check for non local effect, same pair of input modes are again accessed and identical operations, $B(\theta)$ and $B(\delta)$ are performed.  As shown in the schematic, for an ideal beam splitter, we will see the change in mode from one pair to the other when it passes through the first pair of beam splitters.  This results in interference of modes when the same operations are performed on the pair of modes as it was done the first time.  Therefore, the output along all four modes will be dependent on both, $\theta$ and $\delta$.  Mathematically we can write down the action of beam splitter operation in the form,
\begin{align}\label{BS-S1}
|\Psi\rangle_{1-4} &= \begin{bmatrix}
	0 &    B(\theta)  \\
	B(\delta) &  0 
\end{bmatrix} 
\begin{bmatrix}
	0   &    B(\theta) \\
	B(\delta)  &  0
\end{bmatrix}  \Bigg [ \frac{1}{\sqrt 2} (|1000\rangle + |0001\rangle)\Bigg ]  \nonumber \\
& \equiv  \begin{bmatrix}
	B(\theta+ \delta) &    0 \\
	0 & B(\theta+ \delta) 
\end{bmatrix}  \frac{1}{\sqrt 2}  \begin{bmatrix}
	1\\
	0 \\
	0\\
	1
\end{bmatrix}    \nonumber \\
& =  \frac{1}{\sqrt 2}\begin{bmatrix}
	~\cos(\theta + \delta)\\
	i\sin(\theta+ \delta)\\
	i\sin(\theta+\delta)\\
	~\cos(\theta+\delta)
\end{bmatrix}.
\end{align}
From the probabilities of the four output modes of the beam splitter, $E(\theta, \delta)$ can be calculated for various combination of $(\theta, \delta)$, $(\theta, \delta^{\prime})$, $(\theta^{\prime}, \delta)$ and $(\theta^{\prime}, \delta^{\prime})$  and obtain CHSH parameter $S$. From the equivalence relation shown in Eq.\,\eqref{BS-S1} we can also note that the action of independent operation $B(\theta)$ and $B(\delta)$ on pair of input modes will give a same output as operations $B(\theta)$  in series with $B(\delta)$  on the same modes. Therefore, we can say that the local and independent beam splitter operation on multiple modes of single photon is equivalent to combined action of them on all modes. Effectively, the operations will affect the photons in a same way in both case. This is very useful to simplify the experimental  setup to calculate CHSH parameter.

%=================================================================
\begin{figure}[h!]
  \begin{center}
  \includegraphics[width=0.48\textwidth]{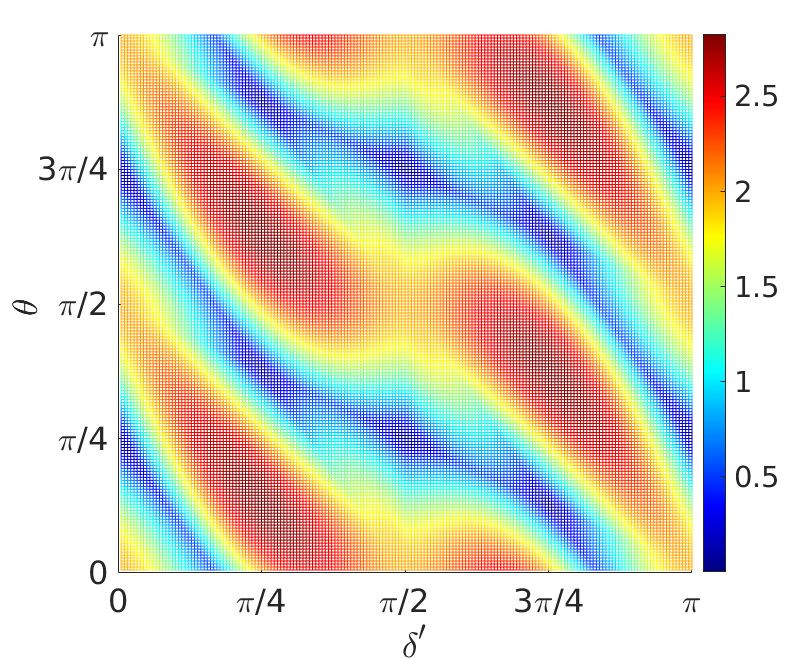}
  \end{center}
\caption{CHSH parameter S as a function of $\theta$ and $\delta^\prime$ in path-encoded single photon when $\delta = 0$ and $\theta^\prime = (\theta + \delta^{\prime})$. Maximum violation of CHSH inequality  can be seen for multiple combination of parameters and shows $S(\theta, \delta) > 2$ for a range of combination of $\theta$ and $\delta^{\prime}$ in a four mode beam splitter setting.} 
  \label{CHSSbeamsplitter}
\end{figure}
%==========================================================
The expression to calculate CHSH parameter in photon occupancy representation will be same as Eq.\,\eqref{bellpara}. In spatial mode setting the effective beam splitter operation will result in,  $E(\theta, \delta) = E(\theta + \delta)$  and $S$ will be,
\begin{align}
S = \lvert E (\theta + \delta) - E(\theta + \delta^{\prime}) \rvert  + \lvert E (\theta^{\prime} + \delta) + E(\theta^{\prime} + \delta^{\prime}) \rvert .
\end{align}
In Fig.\,\ref{CHSSbeamsplitter} we show the CHSH parameter as function of $\theta$ and $\delta^{\prime}$ when $\delta = 0$ and $\theta^{\prime} = (\theta + \delta^{\prime})$.  Using the probabilities, $E(\theta) = \cos (2\theta)$  and S parameter will be in the form, 
\begin{align}
S = \lvert E (\theta) - E(\theta+\delta^{\prime}) \rvert  + \lvert E (\theta + \delta^{\prime}) + E(\theta + 2\delta^{\prime}) \rvert.
\end{align}
We see maximum violation of inequality, $S=2\sqrt{2}$ when $(\theta, \delta^{\prime})$ is $(\pi/8, \pi/4)$, $(5\pi/8, \pi/4)$, $(3\pi/8, 3\pi/4)$ and $(7\pi/8, 3\pi/4)$.   For several other combination of parameters the value of  $S(\theta, \delta ) > 2$ is seen.  For any initial state that is different from  $|\Psi\rangle_{1-2}$ with different splitting ratio,  one can use various combination of beam splitters along the two modes and show the violation of inequality if the photons are entangled in path degree of freedom. 

\begin{figure}[h!]
\includegraphics[width=0.48\textwidth]{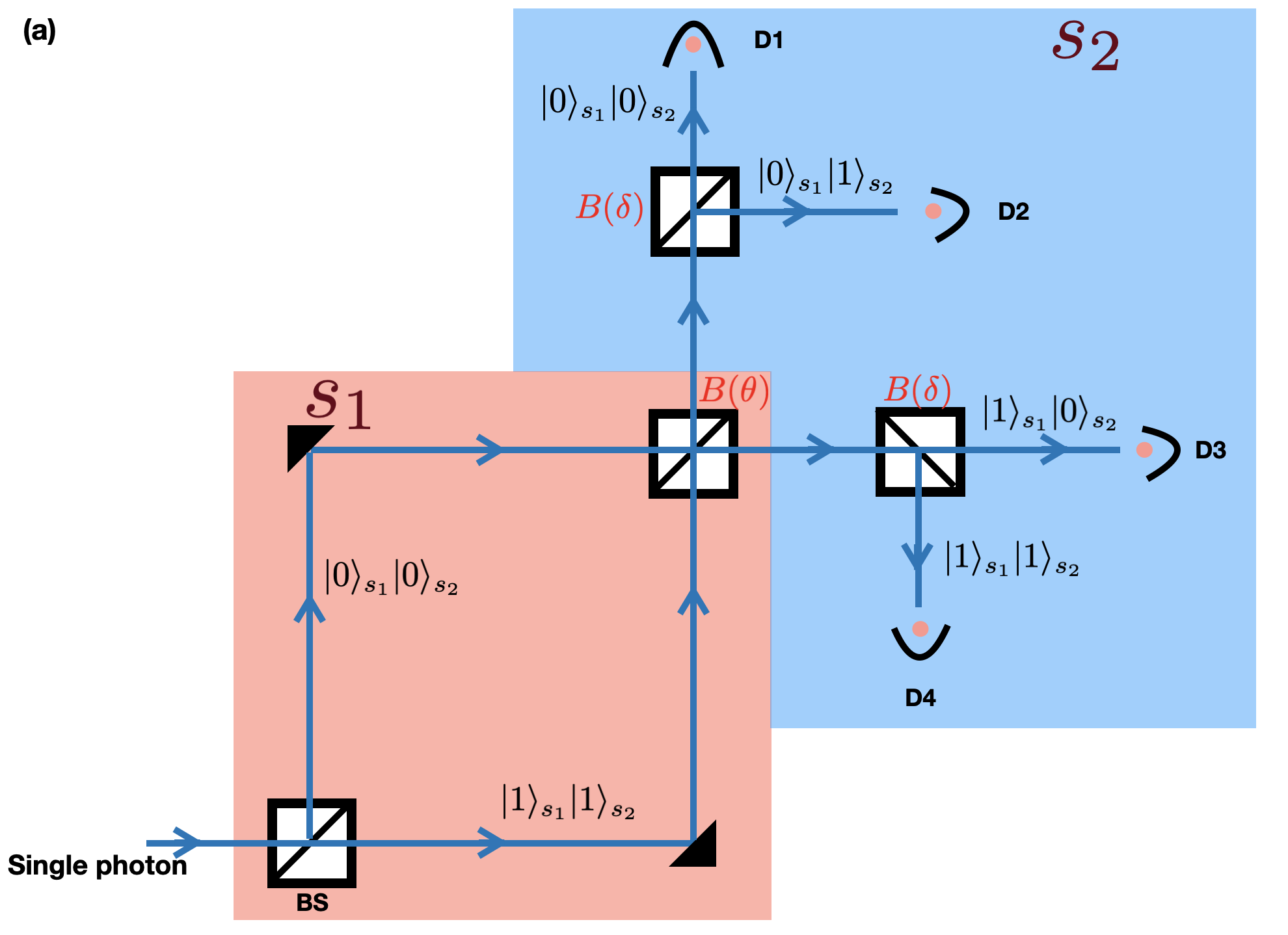}
  \includegraphics[width=0.48\textwidth]{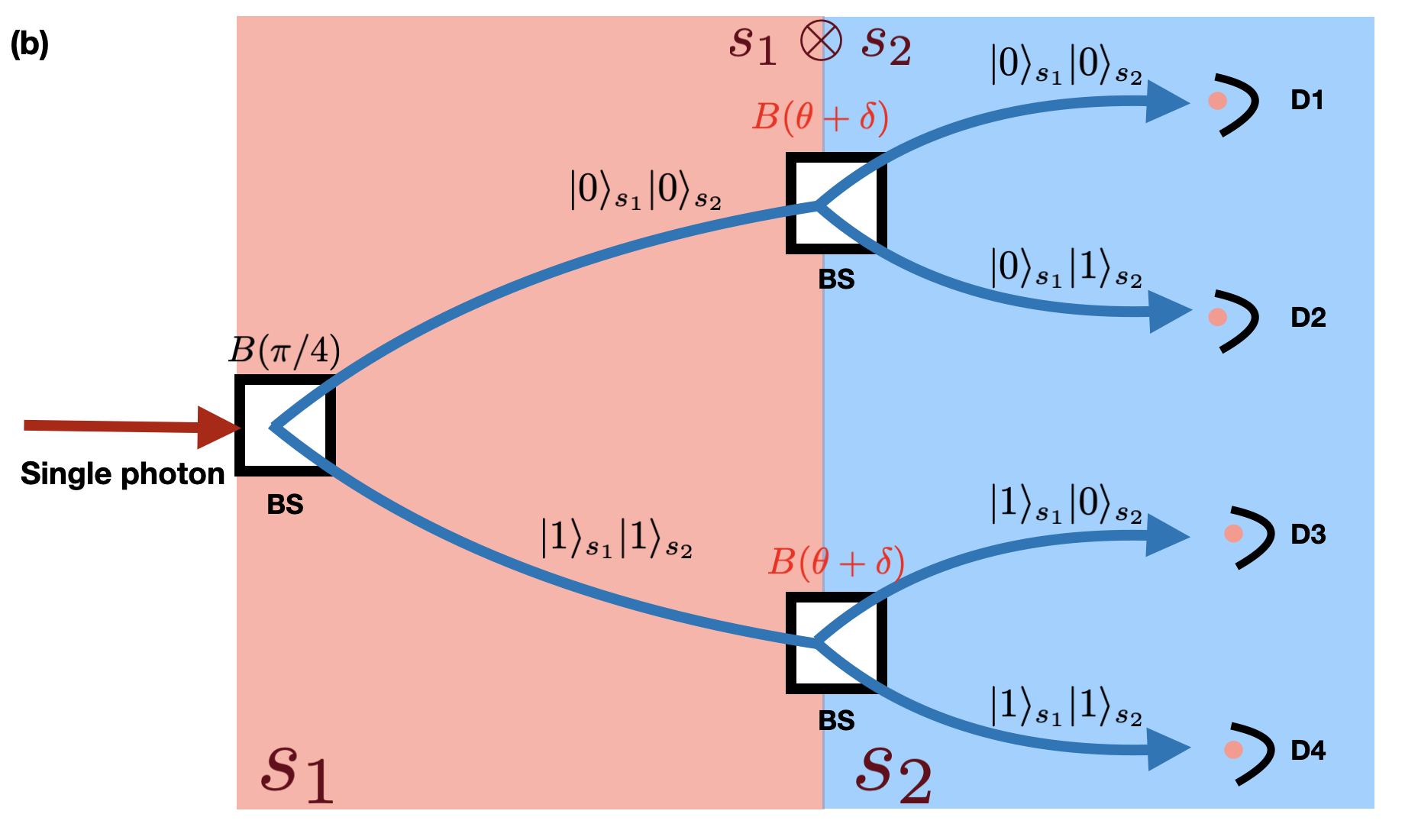}
    \caption{Schematic of combination of beam splitter in two spatial mode representation which is identical to photon occupancy representation. In (a) operators $B(\theta)$ and $B(\delta)$ are independent and photon interfere before they split again and in (b) the operations are applied in combination and paths do not interfere. However, both schemes span over all four basis states and obtain probabilities to calculate CHSH parameter identically.}  
  \label{Schematic1}
\end{figure}

\noindent
{\it Bell's test on path-entangled photon using two spatial mode setting :} The four basis states from path-entangled single photon can also be re-written by associating them with the two spatial modes, $s_1$ and $s_2$ where the basis states are composed of the {\it transmitted} and {\it reflected} component of the photon using states $|0\rangle_{s_i}$ and $|1\rangle_{s_i}$, respectively.  In this representation all four basis states $\{ |0\rangle_{s_1}|0\rangle_{s_2} ,  |0\rangle_{s_1}|1\rangle_{s_2} ,  |1\rangle_{s_1}|0\rangle_{s_2} ,  |1\rangle_{s_1}|1\rangle_{s_2} \}$ can be spanned by introducing single photon in superposition of two input modes of the beam splitter by controlling the transmitted and reflected component of the photons state. In two spatial mode representation, the state with all four basis states will be in the form,
\begin{align}
\label{fockstate1}
|\Psi\rangle_{s_1-s_2}  = a |0\rangle_{s_1}|0\rangle_{s_2} + b |0\rangle_{s_1}|1\rangle_{s_2} + c |1\rangle_{s_1}|0\rangle_{s_2} + d  |1\rangle_{s_1}|1\rangle_{s_2}.
 \end{align}
In Fig.\,\ref{Schematic1}\tr{(a)}  we show the combination of beam splitters  to span over all four basis state in Hilbert space $s_1\otimes s_2$ representation. First, the transmitted and reflected components from mode $s_1$ are spatially separated and they are made to interfere at beam splitter $B(\theta)$. After they interfere it is again subjected to splitting using two beam splitters $B(\delta)$ which represents operation on spatial mode $s_2$.  
%=================================================================
%===============================================================
In Fig.\,\ref{Schematic1}\tr{(b)}, we show the equivalent form of beam splitter setting where photon paths do no interfere but result in an identical outcome when the operation of $\theta$ and $\delta$ are combined together. When second pair of beam splitter is absent, the state $|\Psi\rangle_{s_1-s_2}$ can be written as  
\begin{align}\label{path-photon}
|\Psi\rangle_{p-s}  =  \frac{1}{\sqrt 2}\Big [ |0\rangle_{s_1} |0\rangle_{s_2}  + |1\rangle_{s_1} |1\rangle_{s_2} \Big ]
\end{align}
when first BS$=B(\pi/4)$.

  %=================================================================
Mathematically, using transmitted and reflected components along each path, the action of $B(\theta)$ and $B(\delta)$  will be in the form,   
 \begin{align}\label{BS-S2}
|\Psi\rangle_{s_1-s_2} = & \bigg( B(\theta  )\otimes  B( \delta ) \bigg) |\Psi\rangle_{p-s}   \nonumber \\
= \bigg( B(\theta + \delta )&\otimes \mathbb{I} \bigg) |\Psi\rangle_{p-s}   = \frac{1}{\sqrt{2}}
\begin{bmatrix}
	~~\cos(\theta + \delta)\\
	~i\sin(\theta+ \delta)\\
	~i\sin(\theta+\delta)\\
	~~\cos(\theta+\delta)
\end{bmatrix}.
\end{align}
We can note that the probabilities associated with basis states in Eq.\,\eqref{BS-S1}  and  Eq.\,\eqref{BS-S2}  are identical, thus, equivalence between the two representation of path-entangled single photon in the beam splitter configuration can be established. In general, since we only have access to change the splitting components along paths in both, four spatial mode description and two spatial mode description, one can choose the combination of operations that result in an identical output showing the equivalence, $|\Psi\rangle_{1-4} \equiv |\Psi\rangle_{s_1-s_2}$.   
Therefore, path-entangled single photon in beam splitter setting for all practical purpose can be written in the equivalent form, 
\begin{align}\label{path-photon1}
|\Psi \rangle_{1-2}\equiv |\Psi\rangle_{p-s}.
\end{align}
Therefore, without loss of generality, the CHSH parameter for path-entangled single photon can be calculated using both the configuration and without making paths interfere.

\noindent
{\it Bell's test on path-entangled single photon using polarization degree of freedom :} In general, the  beam splitter setting gives limited access to perform Bell's test on states using controllable splitting ratio between the paths.  To perform Bell's test with more control over splitting ratio, we can use polarization degree of freedom where polarization rotator and polarization beam splitter (PBS) (which is also referred as variable beam splitter) will give full control and serve as a tool box to control photon splitting ratio along the paths.  The scheme for generation of a path-entangled single photon and perform Bell's test  using  polarization degree of freedom will be identical to Fig.\,\ref{Schematic1}. The splitting ratio is controlled using the polarization rotator  (or half-wave plate (HWP)) and PBS. The basis states described earlier in this section, $\{ |0\rangle_{s_1} , |1\rangle_{s_1} \}$ will be replaced with the basis states for polarization degree of freedom, $|H\rangle$ and $|V\rangle$. The rotation operation on polarization degree of freedom is given by 
\begin{align}
R(\theta) = 
 \begin{bmatrix}
	\cos(\theta) & - \sin(\theta) \\
	\sin(\theta) & ~~\cos(\theta)
	\end{bmatrix}.
	\end{align}
Rotating the state of photon $|H\rangle$ by angle $\theta =\pi/4$ we obtain $\frac{1}{\sqrt 2} (|H\rangle + |V\rangle )$ and after passing it through PBS we will have a photon in path-entangled state  in the form of $|\Psi\rangle_{p-s}$ as given in Eq.\,\eqref{path-photon}.   To perform Bell's test we will choose two angle of rotations, $\theta$ to act on the Hilbert space associated with $s_1$, photon polarization and $\delta$ on the Hilbert space associated with spatial mode $s_2$. Since the action of operator on one spatial model will also be an action on the other spatial degree of freedom, the effective state will be, 
\begin{align}\label{rotation1}
& |\Psi\rangle_{s_1-s_2} = (R(\delta)\otimes \mathbb{I})(R(\theta)\otimes \mathbb{I}) |\Psi\rangle_{p-s}  \nonumber \\
&~~= [R(\theta + \delta) \otimes \mathbb{I}]|\Psi\rangle_{p-s}  =\frac{1}{\sqrt 2} \begin{bmatrix}
	~~\cos(\theta + \delta)  \\
	~~\sin(\theta + \delta) \\
	-\sin(\theta + \delta) \\
	~~\cos(\theta + \delta)  	\end{bmatrix}.
\end{align}
%===============================================================
\begin{figure}[h!]
 \centering
 \includegraphics[width=0.47\textwidth]{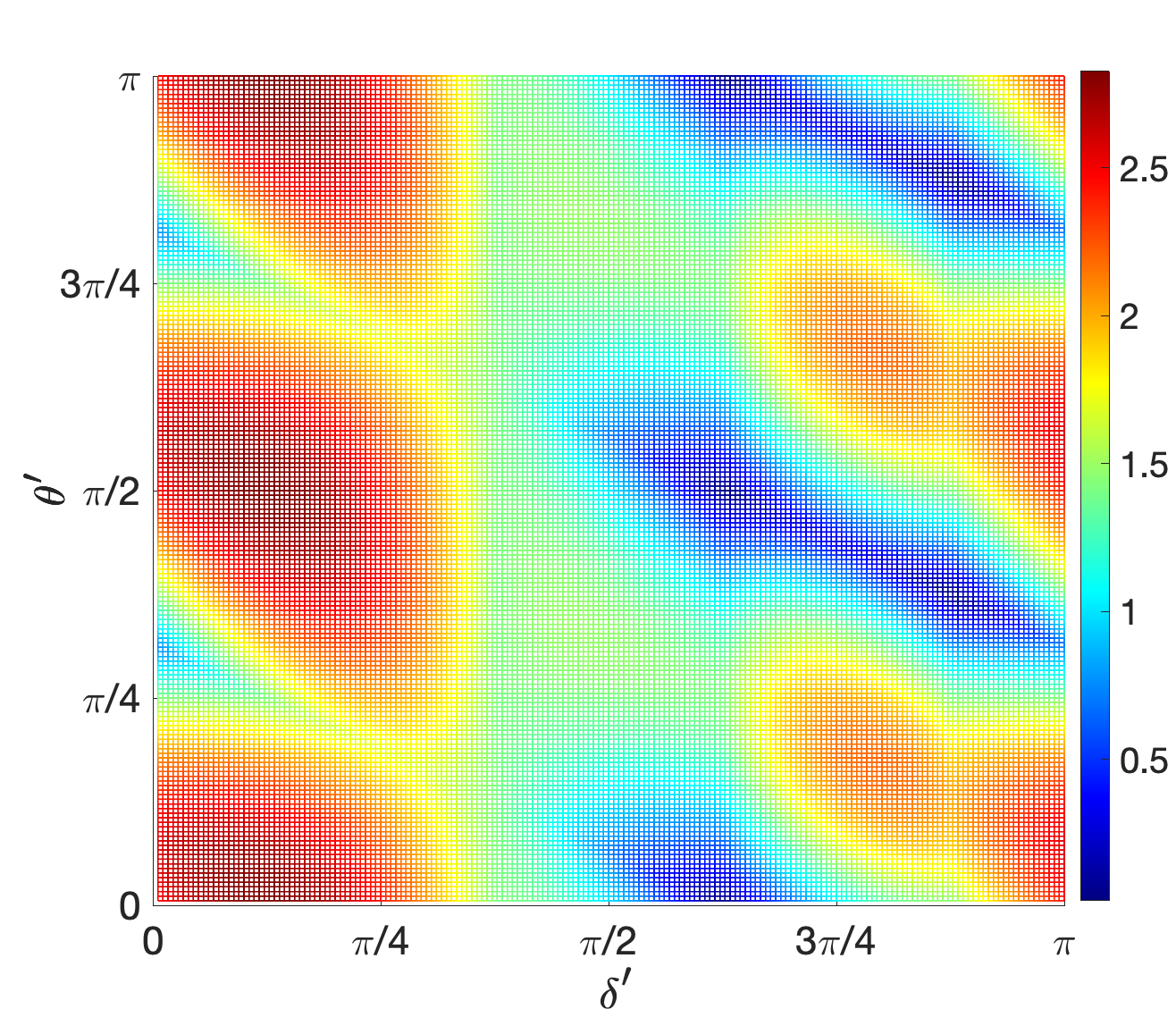}
   \caption{CHSH parameter S as a function of $\theta^{\prime}$ and $\delta^{\prime}$  when $\theta$ and $\delta$ are fixed at $3\pi/4$ and $7\pi/8$. Violation of Bell's inequality ($S(\theta^{\prime}, \delta^{\prime}) > 2$) for various combination of $\theta^{\prime}$ and $\delta^{\prime}$ can be seen.}
\label{Belltest2}
\end{figure}
%================================================================
The expression $E(\theta, \delta)$ which is a composition of probability measurements in different basis state will be 

\begin{align}\label{evalue}
E(\theta, \delta) = P_{00} + P_{11} - P_{01} - P_{10} =  \cos 2(\theta + \delta).
\end{align}
CHSH parameter for different pairs of angles $(\theta, \delta)$ and $(\theta^{\prime}, \delta^{\prime})$ will be,
\begin{align}\label{CHSS}
S(\theta, \delta, \theta^{\prime}, \delta^{\prime}) = \lvert E (\theta, \delta) - E(\theta, \delta^{\prime}) \rvert  + \lvert E (\theta^{\prime}, \delta) + E(\theta^{\prime}, \delta^{\prime}) \rvert.
\end{align}
%Violation of Bell's inequality happen when $S > 2 $. 
In Fig.\,\ref{Belltest2} the parameter $S (\theta^{\prime},  \delta^{\prime})$ obtained when $\theta = 3\pi/4$ and $\delta = 7\pi/8$ is presented.  Violation of CHSH inequality for various combination of parameters and a maximum violation of $S = 2 \sqrt{2}$ is observed for some combination of parameters. 

In this  section we presented three equivalent description for calculating CHSH parameter for path-entangled single photon using four spatial mode, two spatial mode and polarization degree of freedom as the basis states. Though the one using four spatial mode description involves independent and local operations on pair of spatial modes,  the other two use the combined operations in identical form on all input modes.  Since they are equivalent and latter two are simpler for experimental realization, they can be effectively used to calculate CHSH parameter and show violation of Bell's inequality.  We should note that the latter two methods do not ensure local and independent operation on each mode like the four mode description does, but it will still remain a valid method to test path-entanglement of single photons. Below we will show that it can be used as a simple method to test for purity of single photon state by showing the CHSH parameter change in presence of depolorization noise and multi-photon noise.

%In addition to mimicking the path-entangled single photon, using the presence of polarization degree of freedom, a similar setup using the combination of beam splitter and PBS along with polarizer can be explicitly used to calculate CHSH parameter for polarization and path degree of freedom of single photon.   

%========================================
\subsection{Purity test on single photon state}
%========================================

\noindent
{\it Depolarization :} Transition of a pure quantum state to a maximally mixed state can be effectively modeled using quantum depalorizing channel. For the path-entangled single photon state,  it is a linear combination of  Eq.\,\eqref{path-photon}  in its density matrix form $\rho_{p-s} = |\Psi\rangle_{p-s}\langle \Psi|$ and maximally mixed state,
\begin{align}
D_\mathcal{P} (\rho_{p-s}) =  \mathcal{P} \rho_{p-s} + \frac{(1-\mathcal{P})}{2} \mathbb{I}.
\label{depol}
\end{align}
Here $D_{\mathcal{P}}$ is a completely positive trace preserving map and $\mathcal{P}$ is the probability or purity level in this case. By passing the state $\rho^{\prime}_{p-s} = D_{\mathcal{P}}(\rho_{p-s})$ through additional pair of beam splitters, probabilities associated with all four basis states, $E(\theta, \delta)$ and CHSH parameter $S$ can be calculated. \\

\noindent
{\it Multi-photon noise :} One of the common cause for reduction in the purity of single photons is the probability of multi photon pair generation in some of the sources. When multi-photons accompany single photons, purity of single photons reduces and tends towards thermal state. When the photons source comprising of mixture of single photon and muiti-photon states is passed through the beam splitter, the mixed input state can be written in the form, 
\begin{equation}
\rho_{\text{m}} = \mathcal{P} \Big(\rho_{p-s} \Big)  + \frac{(1-\mathcal{P})}{3} \Big(\rho_1  + \rho_2 + \rho_3 \Big).  
\end{equation}
Here, $\rho_{p-s}$ represents the single photons, $\rho_1= \big (\ket{01}+\ket{10} \big) \big(\bra{01}+\bra{10}\big)$ represents multi photon detection in both the detectors, $\rho_2 =  \ket{01}\bra{01}$ and $\rho_3 =  \ket{10}\bra{10}$ represents unresolved multi-photon detection in either of the detector, respectively.  The value of $\mathcal{P}$ purity level of single photons in the mixed input state.  
The path-entangled state for the mixed input state is further passed through a pair of beams splitter to calculate CHSH parameter associated with the input state. 

\begin{figure}[h!]
\centering
\includegraphics[scale = 0.45]{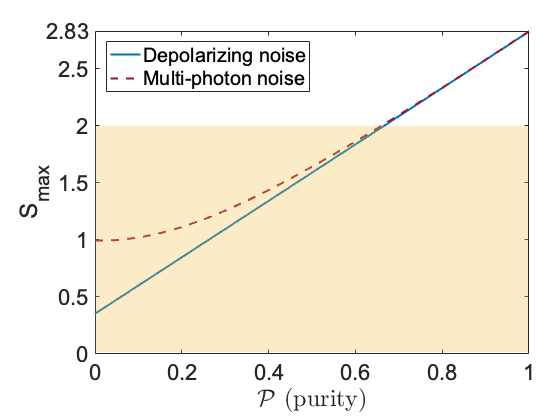}
\caption{Maximum value of $S$ parameter with increase in purity ($\mathcal{P}$) of single photons when subjected to depolorizing noise and when the input state is a mixture of single and multi-photon states. For $\mathcal{P}$ close to $0.7$ we see the transition from validity of CHSH inequality to violation of CHSH inequality.}
\label{theorypurity}
\end{figure}

In Fig.\,\ref{theorypurity} the maximum value for CHSH parameter obtained from numerical calculation with increase  in purity of single photon state is shown. Increase in the value of $S$ and violation of CHSH inequality can be recorded for $\mathcal{P}>0.7$.   

%================================== 
\section{Experimental Method }
\label{Exptsch}
%================================== 
To experimentally demonstrate the violation of CHSH inequality in path-entangled single photon state we will use single photons from SPDC process as source. In SPDC process, the photons are generated in pairs and presence of single photon can be gated by heralding the other photon. If we are not gating the presence of photon by heralding its pair, due to the probabilistic nature of the SPDC process we will have time windows with single photons, no photons or multi-photons. Therefore, when one examines the full temporal behaviour, the  field striking the detector is generally treated as a thermal source. However, instead of examining the full temporal behaviour, if we examine the field and photons detection in the smaller time window we can resolve and record more single photons. Experimentally, with a choice of low pump power, smaller detection time window, better detector efficiency with low dead time and dark counts, one can resolve and detect more single photons even when it is not heralded. Here, we report the experiment performed using both, heralded and un-heralded single photons.  For our experiment with un-heralded source we have chosen low pump power such that the minimum difference between the two detection obtained by averaging over many trial runs is slightly higher than the detector dead time. Keeping the average minimum difference as a time window we resolve a good number of single photons. Though we will have many time window detecting no photons and a very few window detecting multi-photons, single photon and no photon time windows outnumber giving us a random distribution of single photons over time.      

%========================== 
\subsection{Experimental setup}
%==========================
Scheme presented in Fig.\,\ref{Schematic1}\tr{(b)} is considered for the experimental realization. To calculate probabilities needed for CHSH parameter, we need three beam splitter and four single photon detector modules (SPCM). If the single photons are heralded from SPDC, we will need four beam splitter and five SPCM to count the coincidence of photon detection between the heralded photons and each of the four paths. When the single photons are un-heralded we will ideally need only four SPCM to count the number of photons in each path. However, we will show that the same results can be obtained using only two SPCM under repeated measurement with orthogonal setting of HWP and PBS.  For the experimental realization polarization degree of freedom has been used to control the splitting of photons along multiple paths. 
%=================================================================
\begin{figure}[h!]
  \begin{center}
 \includegraphics[width=0.5\textwidth]{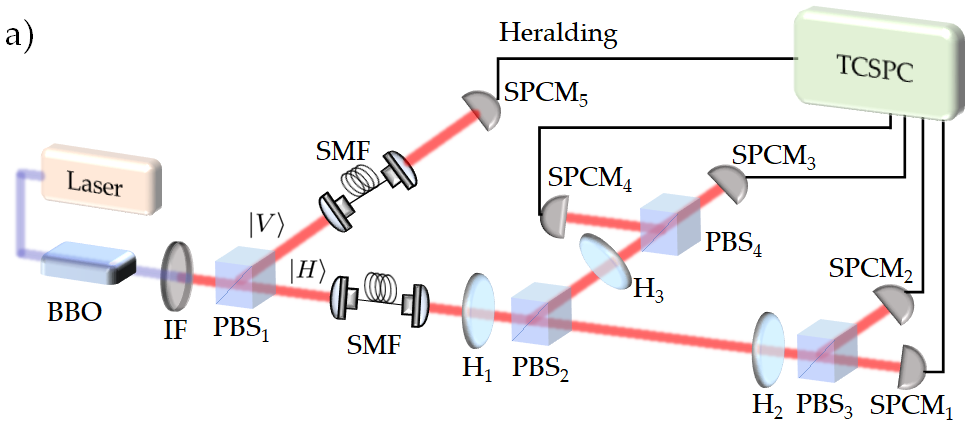}\\
   \includegraphics[width=0.5\textwidth]{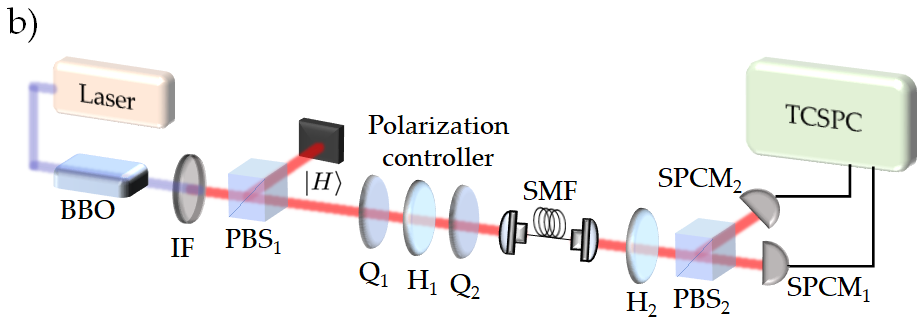}
  \end{center}
\caption{Schematic of the experimental setup using (a) heralded single photon and (b) un-heralded single photon in two detector setting.  A 405 nm laser pumps the BBO nonlinear crystal. The generated orthogonally polarized photon pairs ($|H\rangle$, $|V\rangle$) are separated using a polarization beam splitter (PBS$_{1}$) where $|V\rangle$ is used for heralding (for heralded source) and $|H\rangle$ to generate path-entangled photon. An interference filter (IF) at 810 nm center wavelength with a bandwidth of 10 nm FWHM is used for filtering the pump.  The splitting of photons is controlled using a combination of a half-wave plate and  polarization beam splitter using (a) $H_1$, $H_2$, $H_3$, PBS$_{2}$, PBS$_{3}$, PBS$_{4}$ and for heralded source and (b) $H_2$ and PBS$_{2}$ for un-heralded source with orthogonal setting for second set of measurement. In (b) a polarization controller, consisting of two quarter-wave plates (Q$_{1}$, Q$_{2}$) with a HWP (H$_{1}$) in between, is used to maintain the initial polarization state of single photons to $|H\rangle$. For heralded scheme, (a) photon coincidence measurement are performed using 5 SPCM and for heralded scheme, (b) photon count measurement are done using 2 SPCM.  The output of the SPCM are fed to a time-correlated single photon counter (TCSPC).}
  \label{ExperimentalSetup}
\end{figure}
%==========================================================
The schematic of an experimental setup is shown in  Fig. \ref{ExperimentalSetup}{\tr (a)}. A 1-mm-long BBO nonlinear crystal cut for type-II phase-matching is pumped by a continuous-wave diode laser (Surelock, Coherent) at 405 nm with a pump power.  A half-wave plate (HWP) is placed to control the pump polarization for optimal phase-matching (not shown). A plano-convex lens of f=200 mm (not shown) is used to tightly focus the laser into the crystal. Down-converted degenerate photons at 810 nm are collected using lens of f=30 mm (not shown) and passed through a bandpass interference filter (IF) at centre wavelength 810 nm with bandwidth of 10 nm FWHM to filter input pump. The generated orthogonal polarized photon pairs ($|H\rangle$, $|V\rangle$) are separated using a PBS and then coupled into single-mode fibers (SMF).  From the pair, horizontally ($|H\rangle$) polarized single photons with photon count of 4600 c/s at 40mW are used as our source to generate path-entangled photons and vertically ($|V\rangle$) polarized photon is used as heralding photon.  To generate a path-entangled single photons and perform Bell's test measurements, splitting of photons along different paths are controlled using a combination of HWP and  PBS. The single photon counting measurements along the four paths and the heralding photons  are simultaneously performed using five SPCM ID120 from ID Quantique. By changing the HWP angle from 0 to 360 degrees, we recorded the coincidence count of photons along all the four paths with the heralding photons. The output of the SPCM are fed to a time-correlated single photon counter (Time Tagger, Swabian Instruments). Using the coincidence counts probabilities of all four basis and CHSH parameter $S$ was calculated.

In Fig.\,\ref{ExperimentalSetup}\tr{(b)}, a simplified setup which uses un-heralded single photons as source and only two SPCM to measure the outcome from only one of the basis of the path $|0\rangle_{s1}$ is shown (half of Fig.\,\ref{Schematic1}\tr{(b)}).  With a controlled initial state and series of measurements for different configurations we reproduce the full output expected from the Fig.\,\ref{ExperimentalSetup}\tr{(a)}.  In two detector setting, a polarization controller, consisting of two quarter-wave plates (QWP) with a HWP in between, is used to maintain the initial polarization state of single photons to $|H\rangle$. In our setup, the polarization controller is also used as a depolarizing element to show the transition from validity of CHSH inequality to the violation of inequality with an increase in purity of quantum state of a single photon.

 To test the purity of the single photon state we use two detector setup  and introduce depolarization channel by changing the angles in the waveplates present in polarization compensator.  When the incident angle is not linearly polarized along the symmetry axis of the wave plate, the temporal walk-off will be acquired between the two polarization states. Though the detectors are insensitive to such short temporal walk-offs, the temporal distinguishability acquired during the propagation fulfils the role of the environment in general decoherence models\,\cite{SE11}. Thus, by controlling the angles of the wave plates in polarization compensator, depolarization was introduced to change the visibility of photon state from 100\% to 30\%. Using the number  of photon counts in each SPCM for all angles of HWP ($H_2$) we reconstruct the probability of finding photons in all four basis states of path-entangled single photon and calculate CHSH parameter.

%======================================
\subsection{Measurement procedure}
%======================================
In the experimental setup we have used combination of HWP and  PBS to control the photon splitting along the paths and generate path-entangled state. The HWP operation can be written as
\begin{align}
\label{HWP}
H(\kappa) =  \begin{bmatrix}
	\cos(2\kappa) &  ~~\sin(2 \kappa) \\
	\sin(2 \kappa) &  -\cos(2 \kappa)
	\end{bmatrix}  
\end{align}
and when only probabilities of states are considered, $R(\theta) \equiv H(\kappa /2)$.  To reproduce the probability output identical to the one obtained from Eq.\,\eqref{rotation1} using path-entangled photons from the heralded single photons source, the HWP's $H_2$ and $H_3$ are set to rotate the state by $R(\theta + \delta)$ in different combinations and the coincidence counts that give probabilities of all four basis states are given below,
\begin{align}%\label{prob1}
P_{00} (\theta, \delta)=
  \frac{ C_{1,5}(\theta, \delta) }{\sum_{j=1}^{5}C_{j,5} (\theta, \delta)} ~~;~~
  P_{01} (\theta, \delta)  = \frac{ C_{2,5}(\theta, \delta) }{\sum_{j=1}^{5}C_{j,5} (\theta, \delta)} \nonumber 
\end{align}
\begin{align}%\label{prob3}
P_{10} (\theta, \delta)= \frac{ C_{3,5}(\theta, \delta) }{\sum_{j=1}^{5}C_{j,5} (\theta, \delta)} ~~;~~
P_{11} (\theta, \delta)  = \frac{ C_{4,5}(\theta, \delta) }{\sum_{j=1}^{5}C_{j,5} (\theta, \delta)}.
  \end{align}
$C_{j,5}(\theta, \delta)$ are the number of coincidence detection of photons in SPCM$_j$ and SPCM$_5$.

To reproduce the probability output identical to the one obtained from Eq.\,\eqref{rotation1} using path-entangled single photons from un-heralded single photons in two detector setting, the HWP, $H_2$  is set to rotate the state by $R(\theta + \delta)$ in different combinations and the combination that gives probabilities of all four basis states are given below,
\begin{align}%\label{prob1}
P_{00} (\theta, \delta)=~~~~~~~~~~~~~~~~~~~~~~~~~~~~~~~~~~~~~~~~~~~~~~~ \nonumber \\ 
  \frac{ C_{D1}(\theta, \delta) }{C_{D1} (\theta, \delta) +  C_{D1} (\theta_{\perp}, \delta_{\perp}) +  C_{D2} (\theta, \delta) + C_{D2} (\theta_{\perp}, \delta_{\perp}) } \nonumber 
\end{align}
\begin{align}%\label{prob2}
P_{10} (\theta, \delta)  = ~~~~~~~~~~~~~~~~~~~~~~~~~~~~~~~~~~~~~~~~~~~~~~~ \nonumber \\ 
  \frac{ C_{D2}(\theta, \delta) }{C_{D1} (\theta, \delta) +  C_{D1} (\theta_{\perp}, \delta_{\perp}) +  C_{D2} (\theta, \delta) + C_{D2} (\theta_{\perp}, \delta_{\perp}) } \nonumber 
\end{align}
\begin{align}%\label{prob3}
P_{11} (\theta, \delta)=~~~~~~~~~~~~~~~~~~~~~~~~~~~~~~~~~~~~~~~~~~~~~~~ \nonumber \\ 
  \frac{ C_{D1}(\theta_{\perp}, \delta_{\perp}) }{C_{D1} (\theta, \delta) +  C_{D1} (\theta_{\perp}, \delta_{\perp}) +  C_{D2} (\theta, \delta) + C_{D2} (\theta_{\perp}, \delta_{\perp}) } \nonumber 
\end{align}
\begin{align}\label{prob4}
P_{01} (\theta, \delta)  = ~~~~~~~~~~~~~~~~~~~~~~~~~~~~~~~~~~~~~~~~~~~~~~~ \nonumber \\ 
  \frac{ C_{D2}(\theta_{\perp}, \delta_{\perp}) }{C_{D1} (\theta, \delta) +  C_{D1} (\theta_{\perp}, \delta_{\perp}) +  C_{D2} (\theta, \delta) + C_{D2} (\theta_{\perp}, \delta_{\perp}) }.
\end{align}
$C_{D1}(\theta, \delta)$ and $C_{D2}(\theta, \delta)$ are the first set of number of photon counts in detector 1 and detector 2 when the rotation angles are $\theta$ and $\delta$. 
$C_{D1}(\theta_{\perp}, \delta_{\perp})$ and $C_{D2}(\theta_{\perp}, \delta_{\perp})$ are the second set of number of photon counts in detector 1 and detector 2  when the rotation are $\theta_{\perp}$ and $\delta_{\perp}$. 

\begin{figure}[h!]
  \centering
  \includegraphics[width=0.48\textwidth]{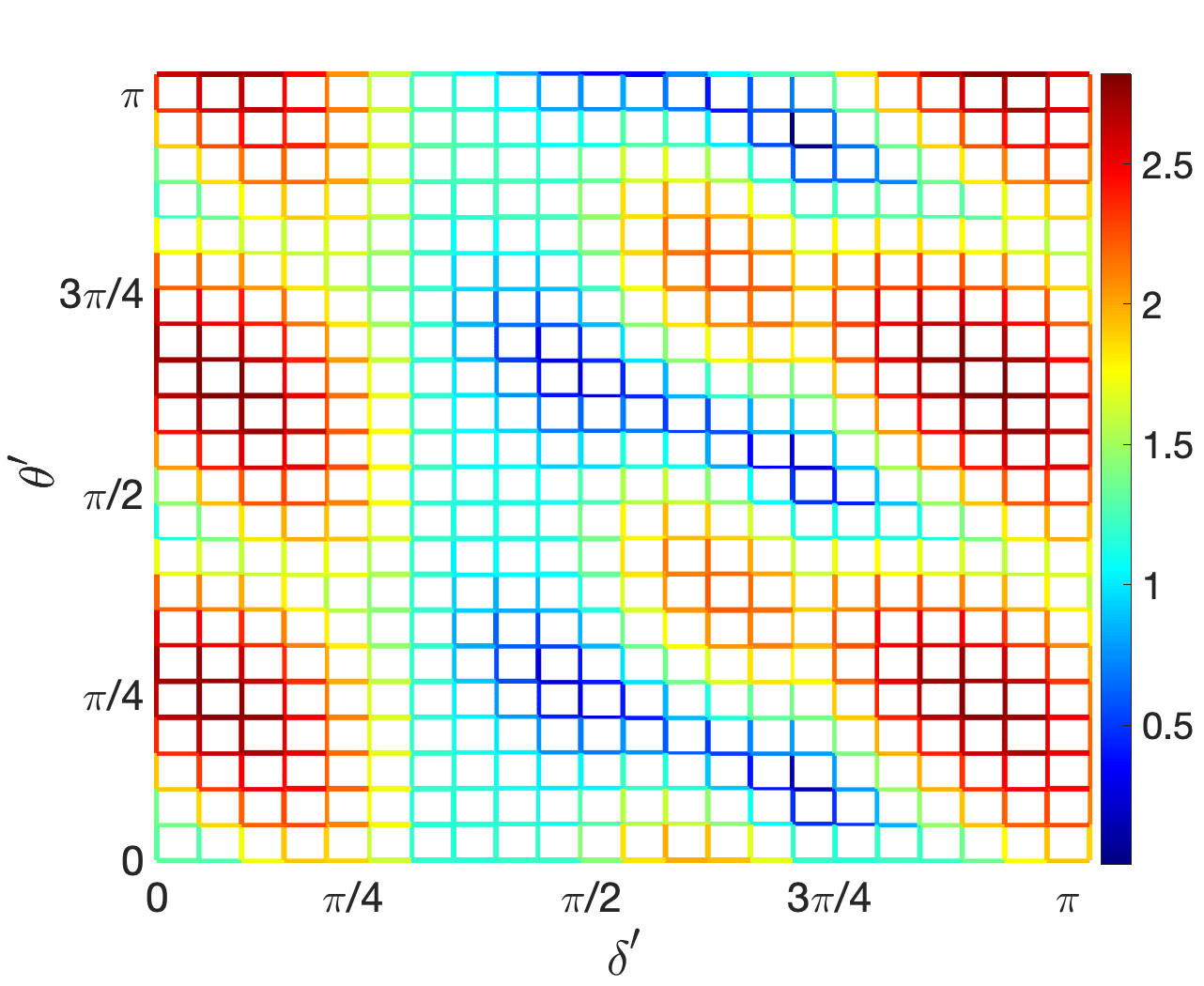}
     \caption{CHSH parameter S as a function of $\theta^{\prime}$ and $\delta^{\prime}$  when $\theta$ and $\delta$ are fixed at $3\pi/4$ and $7\pi/8$. The plot is generated from the experimental photon counts obtained from two photon counting module. Violation of CHSH inequality ($S(\theta^{\prime}, \delta^{\prime}) > 2$) for various combination of values of  $\theta^{\prime}$ and $\delta^{\prime}$ can be seen.}   
  \label{Belltest3}
\end{figure}

For both the experiment the pump power was set to generate 4600 c/s and for heralded single photon source, the coincidence time window was set to 1 ns to calculate $C_{j,5}$.  Almost identical probability values were obtained from both the experiments. Using the probability values,  $E(\theta, \delta)$  CHSH parameter $S$ was calculated.  In Fig.\,\ref{Belltest3},  the value of  $S(\theta^{\prime}, \delta^{\prime})$ when $\theta$ and $\delta$ are fixed at $3\pi/4$ and $7\pi/8$
 using the experimental data for number of photon counts detected per second averaged over 100 trials  using two detector setting is shown.  An identical output (not shown) is obtained path-entangled photons from heralded photon source.  The output presented  in Fig.\,\ref{Belltest3} is in agreement with the numerical simulation obtained for same parameters shown in Fig.\,\ref{Belltest2}. Using the data for full range of rotation of input state, we can reconstruct parameter $S$ for any combination of $\theta$, $\delta$, $\theta^{\prime}$ and $\delta^{\prime}$. The maximum value we could obtain from the experimental data is, $S = 2.82$.
%============================================================
%\subsection{Experimental results}
%================================================

Due to simplicity of two detector setup with minimum devices (optical components and detectors), the deviation from expected maximum value is very low. 

%============================================================
\begin{figure}[h!]
  \centering
  \includegraphics[width=0.235\textwidth]{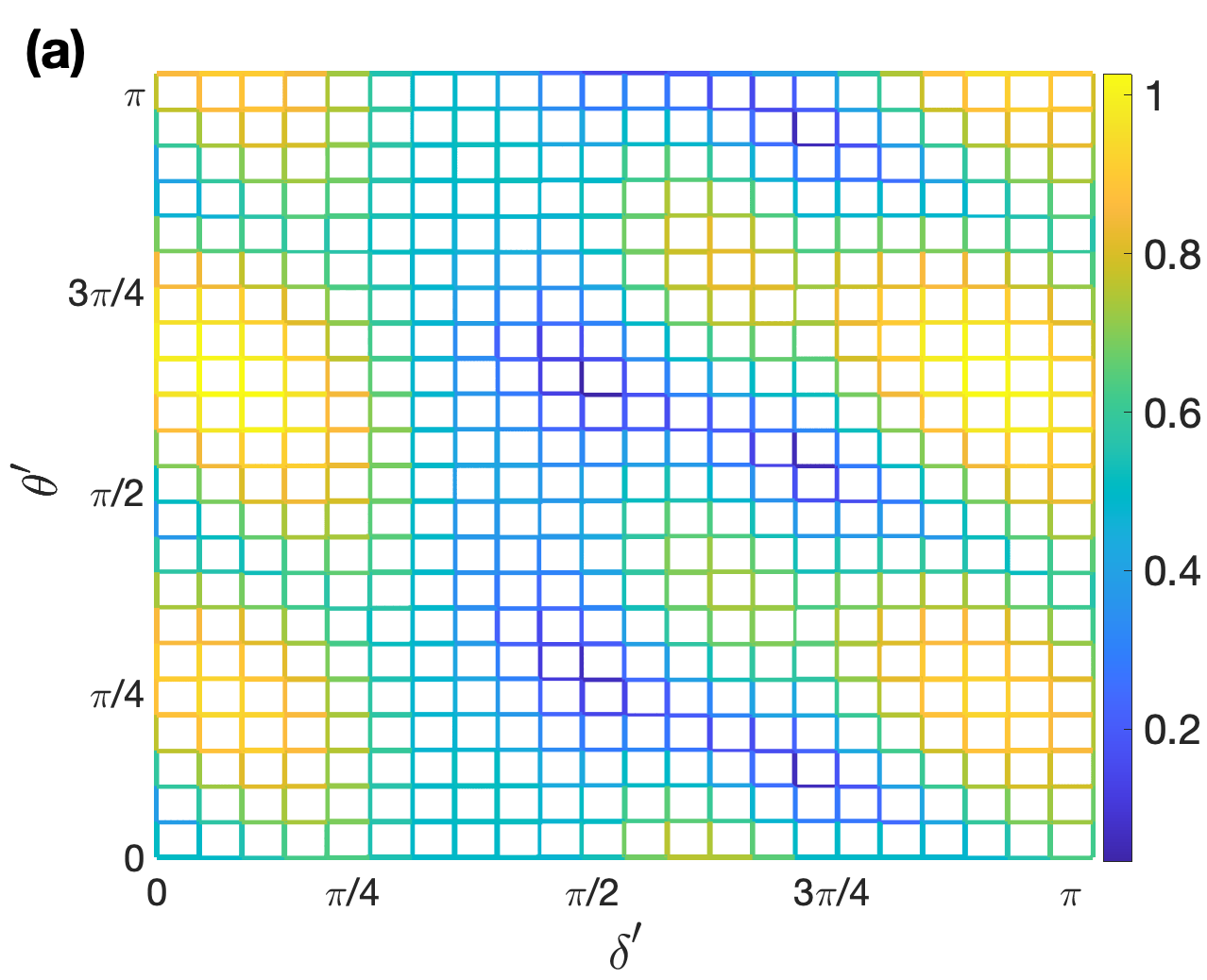}
  \includegraphics[width=0.235\textwidth]{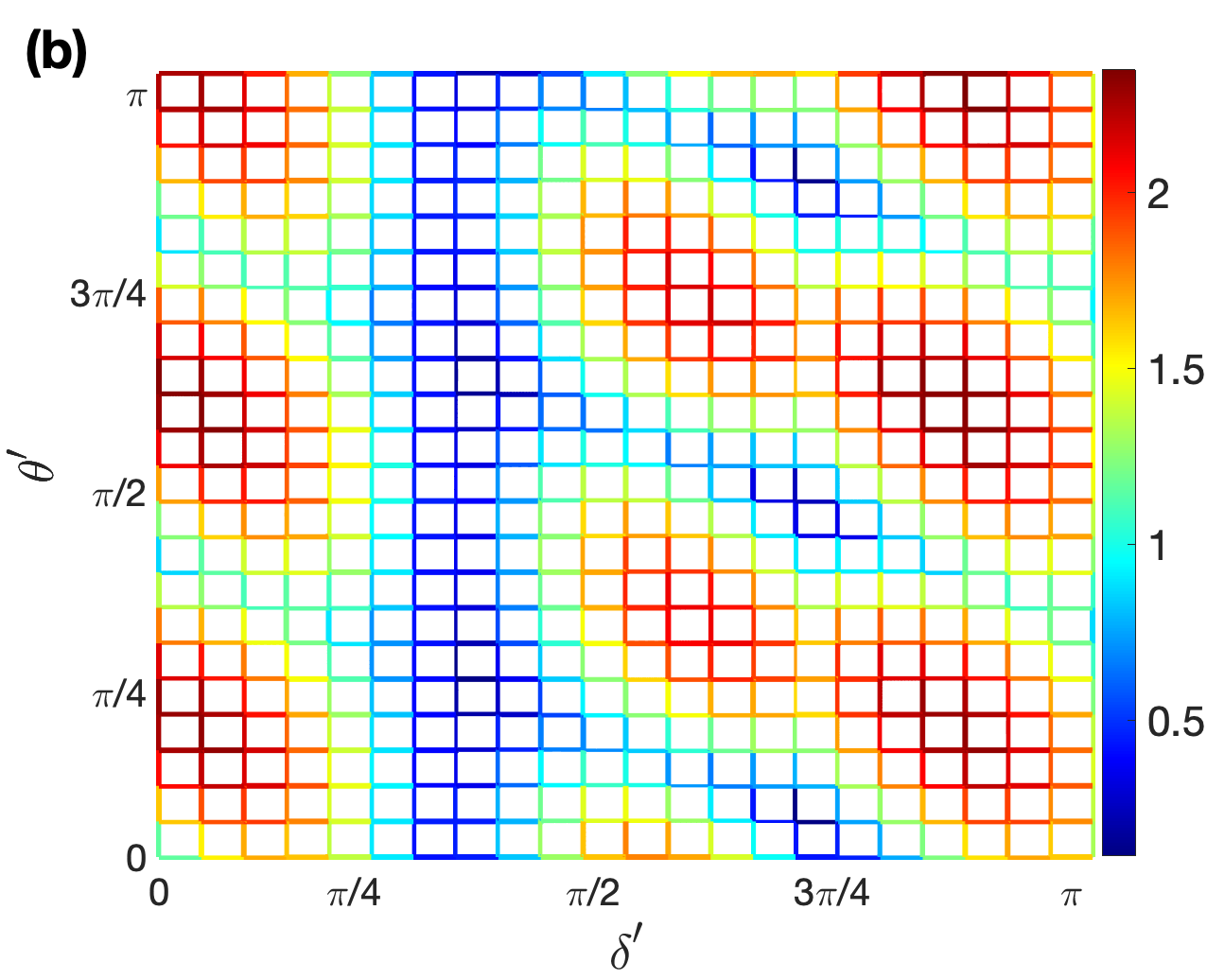}
     \caption{CHSH parameter S as a function of $\theta^{\prime}$ and $\delta^{\prime}$ when $\theta$ and $\delta$ are fixed at $3\pi/4$ and $7\pi/8$. The plot is generated from the experimental photon counts (a) when visibility is 30\% and (b) when visibility is 80\%.   Violation of inequality ($S(\theta^{\prime}, \delta^{\prime}) > 2$ ) for a range of values $\theta^{\prime}$ and $\delta^{\prime}$ can be seen when the visibility is 80\% with maximum value of 2.35 and at 30\% visibility, inequality is not violated.}   
  \label{Belltest4}
\end{figure}
%============================================================
\subsection{Detection and nonlocality loophole}

  In the performed two experiments, we have used coincidence detection for one and the photon count for the other. Though non-coinciding photon detected are ignored in the first case, no post selection is performed when number of photon counts /s are used. Even though the detectors have some inefficiency, they symmetrically affect the photons number counts and hence the measurement does not face a detection loophole. However, thought the local and independent operations are chosen to control pair of beam splitter modes as described in theoretical description, their equivalence with combined operation on all modes does not settle the nonlocality loophole. This work can lead to further investigations towards such nonlocality loophole free Bell's inequality\,\cite{HBD15} in path-entangled or spatial mode setting which are still being debated either through quantum steering approach\,\cite{Ruz21}, in temporal order\,\cite{ZCP19} or in quantum network configuration\,\cite{AKB22}.

%==========================================

\begin{figure}[h!]
  \centering
  \includegraphics[width=0.48\textwidth]{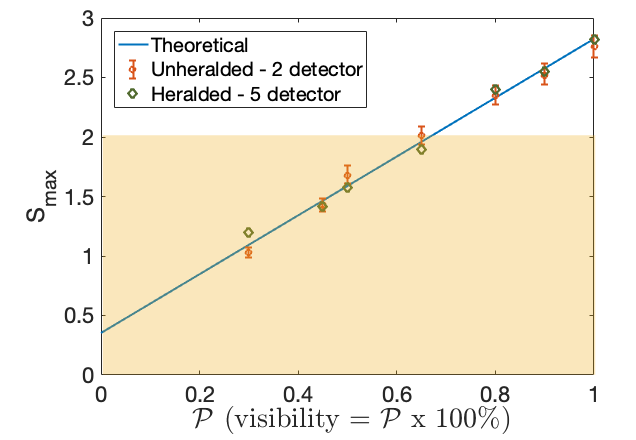}
     \caption{Transition towards violation of CHSH inequality with increase in visibility. Theoretical expectation and experimental values for the maximum value of $S$ when $\theta = 3\pi/4$ and $\delta = 7\pi/4$.}  
  \label{BelltestVisibility}
\end{figure}
\begin{figure}[h!]
  \centering
  \includegraphics[width=0.48\textwidth]{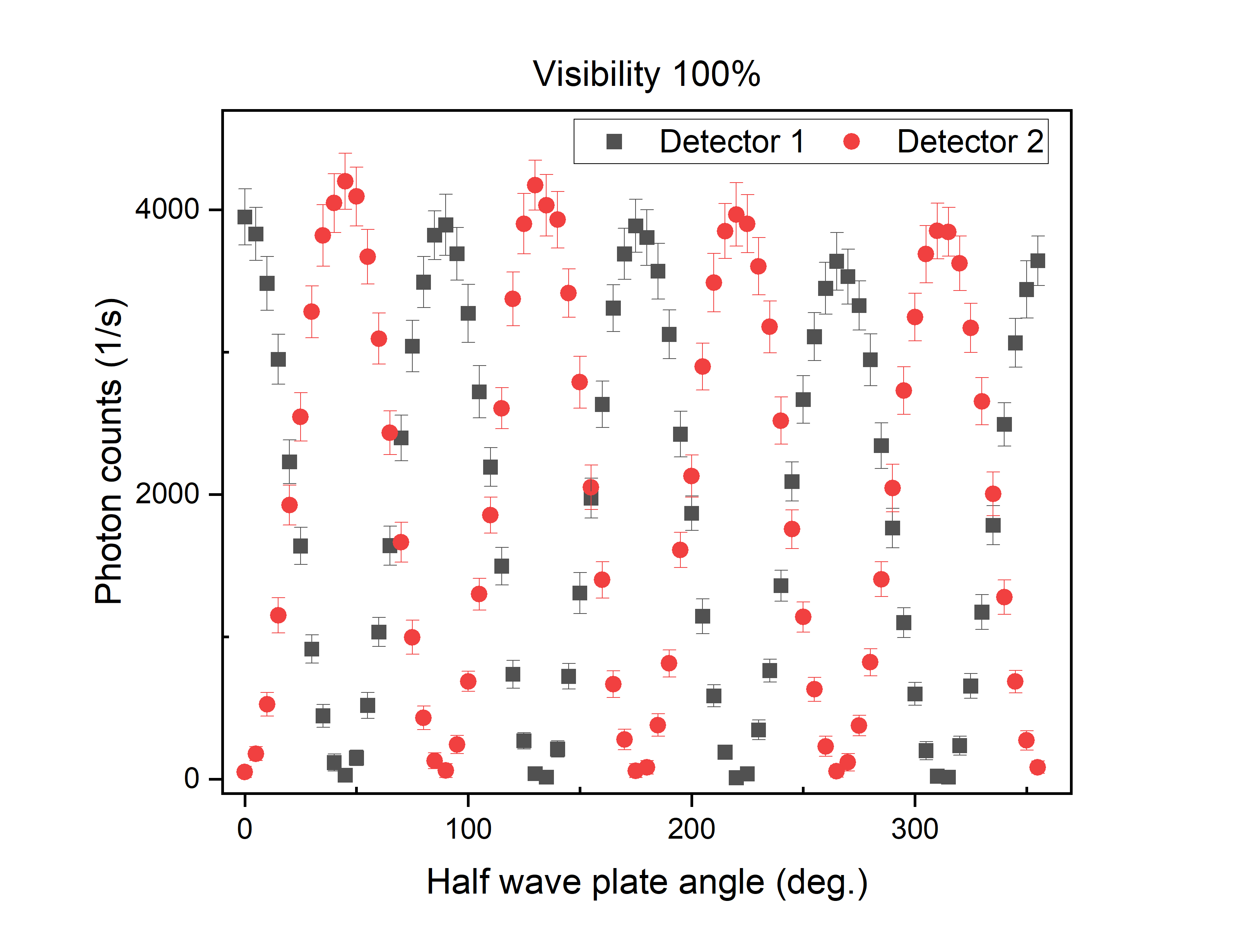}
  \includegraphics[width=0.48\textwidth]{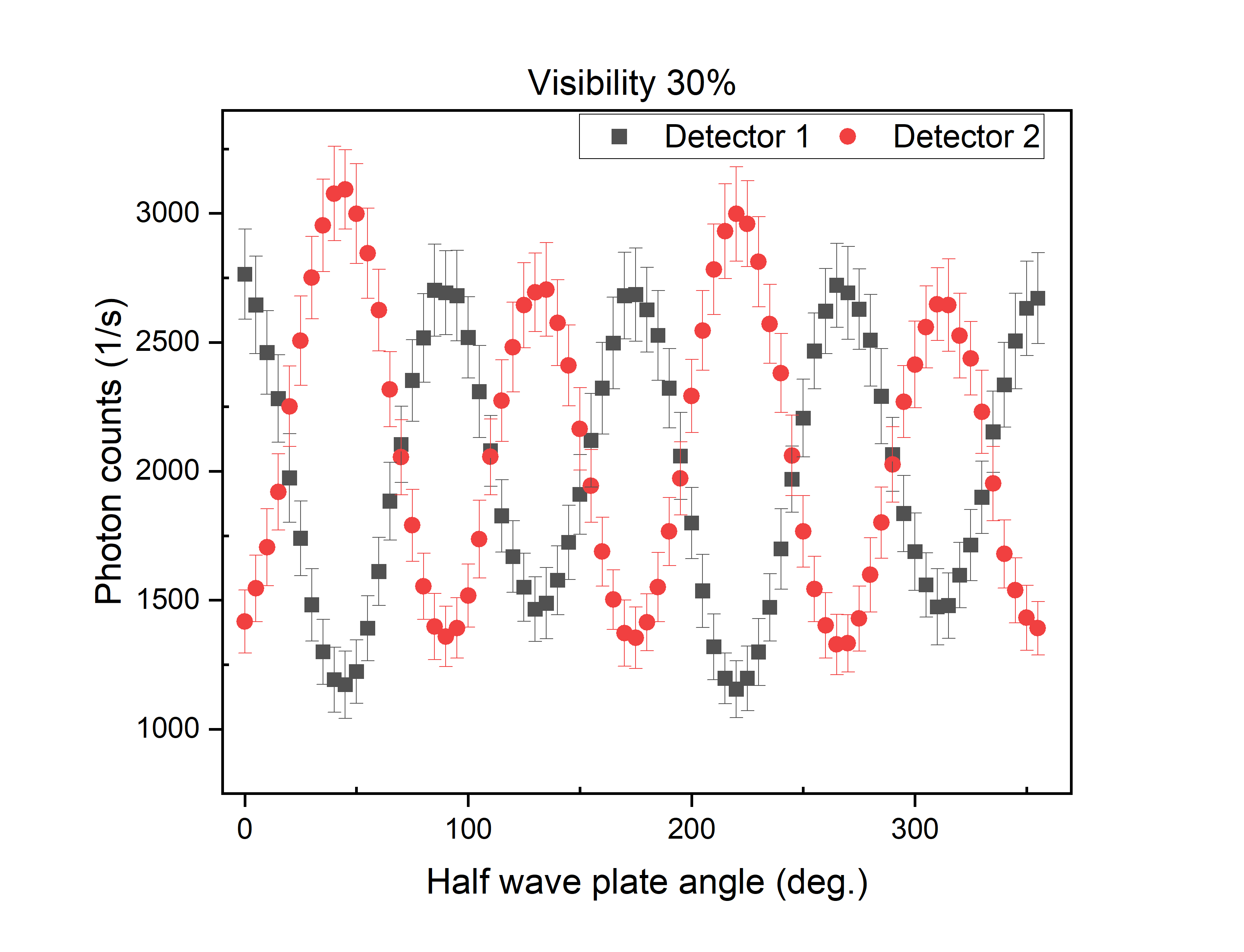}\\
   \caption{Photon counts recorded in detector 1 and detector 2 with change in HWP angle when the visibility is 100\% (pure state) and when the visibility is 30\% (mixed state).  Photons counts/ second was averaged over 100 trials.}  
  \label{photoncount_visibility}
\end{figure}
%==============================================================================

\subsection{Test for purity of single photon state}

%============================================================
Depolarizing channel with controllable parameters on single photon can be effectively realized using birefringent crystals\,\cite{SE11}. In path-entangled single photon, transition from a pure state to a mixed quantum state can be seen with decrease in visibility of quantum state. Using the polarization compensator, QWP-HWP-QWP combination in the setup one can control the visibility (purity) of the quantum state by changing the angles in the polarization compensator. In Fig.\,\ref{Belltest4} we show the value of CHSH parameter $S$ from the experimental data for a set of values of $\theta$ and $\delta$ when visibilities are 30\% and 80\%, respectively. Though the pattern of change in $S$ remains same for both visibility values, we can see a significant difference in the value of $S$ which is above $2$ for 80\% visibility and below $1$ for 30\% visibility. Fig.\,\ref{BelltestVisibility} illustrates the  experimentally obtained value of $S$ for different percentage of visibility along with the theoretical values of $S$ as function of $\mathcal{P}$ by introducing depolarizing channel as shown in Eq.\,\eqref{depol}. With an increase in purity $\mathcal{P}$, we clearly see the transition towards violation of CHSH inequality.  For both experimental setting, with heralded coincidence detection and un-heralded with photon count in two detector setting, we see the experimental value of $S$ being quite close to theoretical value.  In Fig.\,\ref{photoncount_visibility} we have provided the photon counts we recorded in two detectors setting with change in HWP angle when visibility of photon state was 100\% and 30\%. 

%===============================================
\section{conclusion}
\label{conc}
%===============================================

In this work we have presented a theoretical framework to calculate CHSH parameter for path-entangled single photons in beam splitter setting using local and independent operation on sub-set of four spatial modes of the beam splitter. Using its equivalence relation it has been experimentally demonstrated. By experimentally mimicking the effect of depolarizing channel, we have validated the use of CHSH as a test of purity for single photon state. The theoretical results and experimental results we have obtained are in good agreement with each other. Though our theoretical scheme proposes the use of four detector (SPCM) module with local and independent operation on spatial modes which involves interference of paths, using the equivalence relation shown, we have used combined operations on photon state which does not involve any interference of paths to calculate CHSH parameter. We have further demonstrated that by using an alternate measurement procedure accessing only pair of basis state each time, a simpler two detector module can be used to calculate CHSH parameter. The maximum value of CHSH parameter we have obtained is $S=2.82$ and we have observed the transition towards violation of CHSH inequality happening around purity value of $\mathcal{P} = 0.7$ and above. In our  procedure, polarization degree of freedom was only used to have control over the splitting of photon along different paths and was not explicitly used as one of degree of freedom to calculate CHSH parameter but same procedure can be extended to calculate CHSH parameter for polarization and path degree of freedom. The simple non-interferometric scheme we have used for the demonstration makes it a practically efficient way to test purity of single photon state from any source.  The measurement procedure we have adopted will be very useful to experimentally calculate quantum correlation in particle-path (spatial mode) entangled system.

%============================================================

\vskip 0.2in

{\bf Acknowledgment: } We acknowledge the support from the Office of Principal Scientific Advisor to Government of India, project no. Prn.SA/QSim/2020.

%========================================================

\end{document}